\documentclass[article,twocolumn]{revtex4}
\usepackage{graphicx}
\usepackage{amssymb}
\usepackage{bm}
\usepackage{hyperref}
\usepackage{epstopdf}
\usepackage{color}

\begin{document}

\title{Effects of New Viscosity Model on Cosmological Evolution}

\author{Jiaxin Wang $^{1}$}\email{jxw@mail.nankai.edu.cn}
\author{Xinhe Meng$^{1, 2}$}\email{xhm@nankai.edu.cn}

\affiliation{$^{1}$Department of Physics, Nankai University, Tianjin
300071, P.R.China}
\affiliation{$^{2}$ Kavli
Institute of Theoretical Physics China,\\CAS, Beijing 100190,
China.\vspace{1cm}}
\date{Sep 15th 2013}
%
\begin{abstract}
Bulk viscosity has been intrinsically existing in the observational  cosmos evolution with various effects for different cosmological evolution stages endowed with complicated cosmic media. Normally in the idealized ``standard cosmology"  the physical viscosity effect is often negligent in some extent by assumptions, except for galaxies formation and evolution or the like astro-physics phenomena. Actually we have not fully understood the physical origin and effects of cosmic viscosity, including its functions for the universe evolution in reality. In this present article we extend the concept of temperature-dependent viscosity from classical statistical physics to observational cosmology, especially we examine the cosmological effects with possibility of the existence for two kinds of viscosity forms, which are described by the Chapman's relation and Sutherland's formula respectively. By considering that a modification of standard model with viscosity named as $\Lambda$CDM-V model is constructed, which is supported  by data fitting. In addition to the enhancement to cosmic age value, the $\Lambda$CDM-V model possesses other two pleasing features: the prediction about the no-rip/singularity future and the mechanism of smooth transition from imperfect cosmological models to perfect ones.
\end{abstract}

\maketitle
\section{Introduction}

The perfect fluid cosmological models are built on the dynamics governed by Einstein's theory of general relativity with the cosmological principle valid at large cosmic scales and the assumption of cosmic compositions as idealized perfect fluids, which means that all components of the matter-energy in our universe are considered as perfect fluid without any viscosity which may be interpreted as a result of having taken a roughly gross grain approximation. The well-known $\Lambda$CDM model, which is regarded as the most idealized standard description for current cosmology, obtains a constant dark energy term (the famous cosmology constant) that permeates the space everywhere and globally fits the observational data sets very well~\cite{winberg}. But researches have shown that dynamical dark energy models are also allowed by both observations and theoretical considerations~\cite{wmap9y1,gbz}. So far the related research work in this line has produced a variety of improved models, and one direction is studying the practical effects of the physically existing viscosity on the cosmic evolutions and its manifestations~\cite{xh1,xh2,xh3,example1,example2,example3,example4,example5,turb}. It is a natural fact  that viscosity exists in the universe evolution history with its eminent role in the thermal plasma as from hot Big Bang like stages and the important effect of turbulence viscosity has been observed both in intra-galactic and inter-cluster medium.

Even though the existence of viscosity is not a problem, we are confused by its possible origin in the cosmos and how the cosmic viscosity affects the cosmological evolution. In other words, we are pursuing after physically reasonable explanations and acceptable behaviours of the large-scale existing viscosity in the universe.

Much efforts have been made to study imperfect models with various forms of viscosity term~\cite{example1,example2,example3,example4,example5,schwarz2,JDB1,JDB2,JDB3,JDB4}. We believe that the perfect fluid universe should be an approximation of imperfect fluid universe when it inflates, where the evolutional behaviour of viscosity should be functioning.

According to the observationally homogeneous and isotropic property of the large scale theoretical universe discussed above, we focus on the the bulk viscosity (or the 2nd viscosity) term, while neglecting dissipation and shear viscosity which are incompatible with the cosmological principle.
The full expression of moment-energy tensor containing extra pressure of bulk viscosity reads
\begin{equation}
  T_{\mu\nu}=\rho U_\mu U_\nu+(p-3{\rm H}\zeta)h_{\mu\nu}, \label{stress tensor}
\end{equation}
where $\rho$ is energy density, $p$, the isotropic pressure of perfect fluid, $U^\mu=(1,0,0,0)$, the four-velocity of the cosmic fluid in co-moving coordinates, $h_{\mu\nu}=g_{\mu\nu}+U_\mu U_\nu$ represents the projection tensor, and $3{\rm H}=\partial_\mu U^\mu$.

We will propose an imperfect fluid model with temperature-dependent viscosity origin in the following section. Model constrainted with cosmic observational data-sets are included in the third section. We discuss and perform some numerical contrasts between the viscous model and $\Lambda$CDM model in section four. In the end, we briefly show some features of the new model with conclusions.

\section{viscosity}

According to Eq.~(\ref{stress tensor}), the effective pressure which includes the contribution from bulk viscosity reads
\begin{equation}
  p=w\rho-3{\rm H}\zeta, \label{eq5}
\end{equation}
where $\zeta$ is the bulk viscosity parameter, and $w$ is the EoS parameter of cosmic fluid except for the viscosity. For simplification, we neglect the radiation since it contributes little to the late-stage evolution scenario.

In this article we try to extend the concept which is one possible origin of cosmic viscosity---molecular-like interaction---to cosmological researches.
On large-scale we assume: \\
First, the interaction between galaxies or baryon clumps and other thermal interactions occurred in local area could be effectively depicted by molecular-like interaction seen on large-scale. \\
Second, we try to use linear-evolution law to mimic the evolutional behaviour of an universal temperature which can represent the large-scale effects of those local phenomena.\\
We are quite confident that at the last scattering, such universal temperature is about $3000 K$ and it must be approaching zero corresponding to the dilution of cosmic-fluid through late-stage expansion. As a test, we set the effective temperature as $2.73 K$, equalling to the CMB black-body temperature at present.

In this paper, we adopt two appropriate molecular-like collision/ interaction models for cosmology study, from which the viscosity is physically generated; one is based on the Chapman-Enskog equation~\cite{Champan} for dilute multicomponent gas mixtures~\cite{Joseph} (to a first approximation) which can be simplified as
\begin{equation}
  \zeta=\frac{x_i^2}{A x_i^2+B}(T^{\frac{1}{2}}),\label{chaps}
\end{equation}
where $A$ and $B$ are generalized temperature-independent factors which include collision diameter, collision integral, and molecular mass. $x_i$ represents the fraction of each component.

The other one is the Sutherland's formula~\cite{Champan} which reads
\begin{equation}
  \zeta=\zeta'(\frac{T}{T'})^{\frac{3}{2}}\frac{T'+S}{T+S},\label{suths}
\end{equation}
where $S$ is Sutherland's constant, the prime represents reference values here.

The above two models are both semi-theoretical or empirical with first order approximation. The Chapman-Enskog equation assumes that molecules possess only translational kinetic energy, while Sutherland's formula assumes molecules are smooth rigid elastic spheres surrounded by fields of attractive force. Both forms are available for building cosmological models under proper simplification, since the empirically determined parameters are impossible to be given precisely on cosmic large scales. At low temperature ($T\ll 300K$), a general equation which can approximately represent the main feature of Chapman's and Sutherland's formulae reads
\begin{equation}
  \zeta=\zeta_0T^\alpha, \label{eq:11}
\end{equation}
where $\zeta_0$ is considered as a simplified coefficient, and $\alpha$ equals to $\frac{1}{2}$ or $\frac{3}{2}$, which are chosen specifically in order to physically represent Chapman's or Sutherland's limit formula respectively. Theoretically, $\alpha$ could be lower than $\frac{3}{2}$  in Sutherland's formula (\ref{suths}), we pick this value as the upper limit in late-time cosmological evolution. Thus, the Chapman's model and Sutherland's model of viscosity origin can be unified by Eq.~(\ref{eq:11}) and extended on cosmic large scales in late-stage evolutional era.

\section{cosmological model with constraints}
At first we tried the unified dark fluid model, which unfortunately may lead to singularity phenomenon, indicating that unified imperfect dark fluid model can hardly explain the full richness of dark energy phenomenon, this result is in consistent with previous researches~\cite{JDB1,JDB2,JDB3,JDB4,schwarz1}.
We keep in mind that so far by the global fitting, any feasible cosmology model to include the dominated dark energy should not deviate too much away from the $\Lambda$CDM model, which leads us to move on to the study of effective  viscosity functioning on the basis of $\Lambda$CDM.

\subsubsection{basic equations}

In this subsection, we focus on the application of viscosity to the $\Lambda$CDM model. The expression of viscosity follows Eq.~(\ref{eq:11}) which contains both the physical content of Chapman's equation and Sutherland's formula, while the corresponding Einstein's equations which include the constant dark energy $\Lambda$ term  in the flat FRW metric read
\begin{eqnarray}
  {\rm H}^2=\frac{8\pi G}{3}\rho=\frac{8\pi G}{3}(\rho_m+\rho_\Lambda),\label{eq19}\\
  {\rm H}^2+\dot{\rm H}=-\frac{4\pi G}{3}(\rho+3\frac{p}{c^2}),\label{eq20}
\end{eqnarray}
where $\rho_m$ is energy density of pressure-less matter,  $\rho_\Lambda$ represents energy density of the constant dark energy,  and $c$ is the light speed in the vacuum (in natural units $c$ equals to one). The expression of the effective pressure $p$ which includes the viscosity contribution in Eq.~(\ref{eq20}) then reads
\begin{equation}
  p=-\rho_\Lambda c^2-3{\rm H}\zeta,\label{eq21}
\end{equation}
where the first term is the effect of constant dark energy which EoS parameter $w_{DE}=-1$ exactly, and the second term comes from the effect of viscosity.

The combined form of Eqs.~(\ref{eq19})--(\ref{eq21}) reads
\begin{equation}
  \dot{\rm H}=-\frac{3}{2}\Omega_{m0}{\rm H}_0^2(1+z)^3+12\pi G{\rm H}\zeta,
\end{equation}
which can be parametrized as
\begin{equation}
  \frac{\partial E}{\partial z}=A\frac{(1+z)^2}{E}-\frac{B}{H_0}(1+z)^{\pm\frac{1}{2}},\label{eq23}
\end{equation}
where $A=\frac{3}{2}\Omega_{m0}$ and $B=12\pi G\zeta_0T_0^\alpha$ with $\alpha$ equals to $\frac{1}{2}$ or $\frac{3}{2}$ according to the specific formulae of viscosity. The power index $\pm\frac{1}{2}$ in the above equation corresponds to Sutherland's formula and Chapman's relation respectively. We name this model $\Lambda$CDM-V where V is short for the viscosity. For convenience, letter ``a'' and ``b'' are added as postfixes when $\alpha$ equals to $\frac{1}{2}$ and $\frac{3}{2}$ respectively. The new model differs from the standard $\Lambda$CDM with an additional term $\frac{B}{{\rm H}_0}(1+z)^{\pm\frac{1}{2}}$ in Eq.~(\ref{eq23}), which has no simply analytical solution. We will use numerical method by Bayesian analysis in model constraining. Notice here exist two theoretical restrictions for Eq.~(\ref{eq23}), i.e: $E(z=0)=1$ which is a natural boundary condition, and $B>0$ which must be ensured due to the physical existence of viscosity.

\subsubsection{astrophysical data constraints}

In this subsection we have proposed a model which explicitly announces that the universe consists of baryons, cold dark matter, constant dark energy and viscosity, and both the dark energy and viscosity contributes to the effective pressure. In addition to the observational H(z)~\cite{h1,h2,h3,h4,h5,h6,h7,h8} listed in Table~(\ref{tab0}) and SNe Ia data-sets (Union 2.1) with and without systematic errors, the baryon acoustic oscillation (BAO) data-sets~\cite{bao1,bao2} listed in Table~(\ref{tab00}), which has strong constraints on dark energy models will also be included in model constraining. We should mention that several observation H(z) data points in the general H(z) data catalogue are from BAO. We will not treat them twice in doing the statistic analysis \cite{Du1,Du2} for the models, which means H(z) data from Refs.~\cite{h6,h7,h8} will not be included in joint analysis of BAO and H(z). The method we adopted for SN Ia analysis with systematic errors is introduced in Ref.~\cite{sys-err}, providing more reliable results than analysis without systematic errors.

\begin{figure}[h!]
\begin{center}
\includegraphics[width=3in]{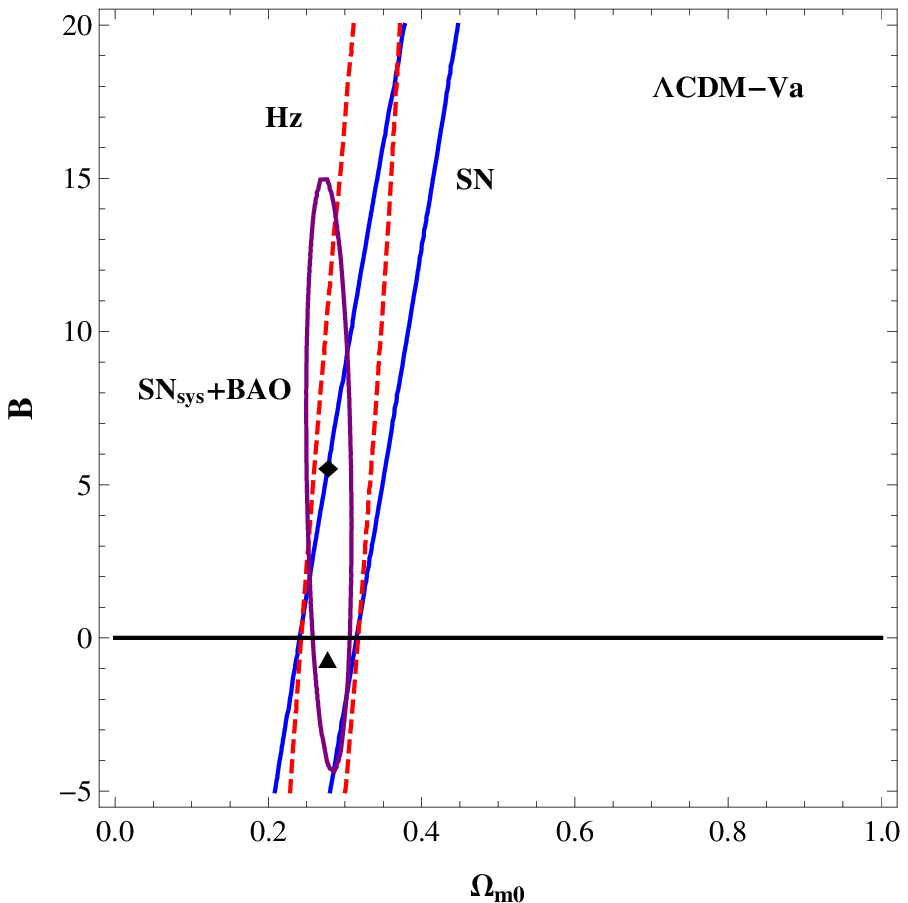}
\includegraphics[width=3in]{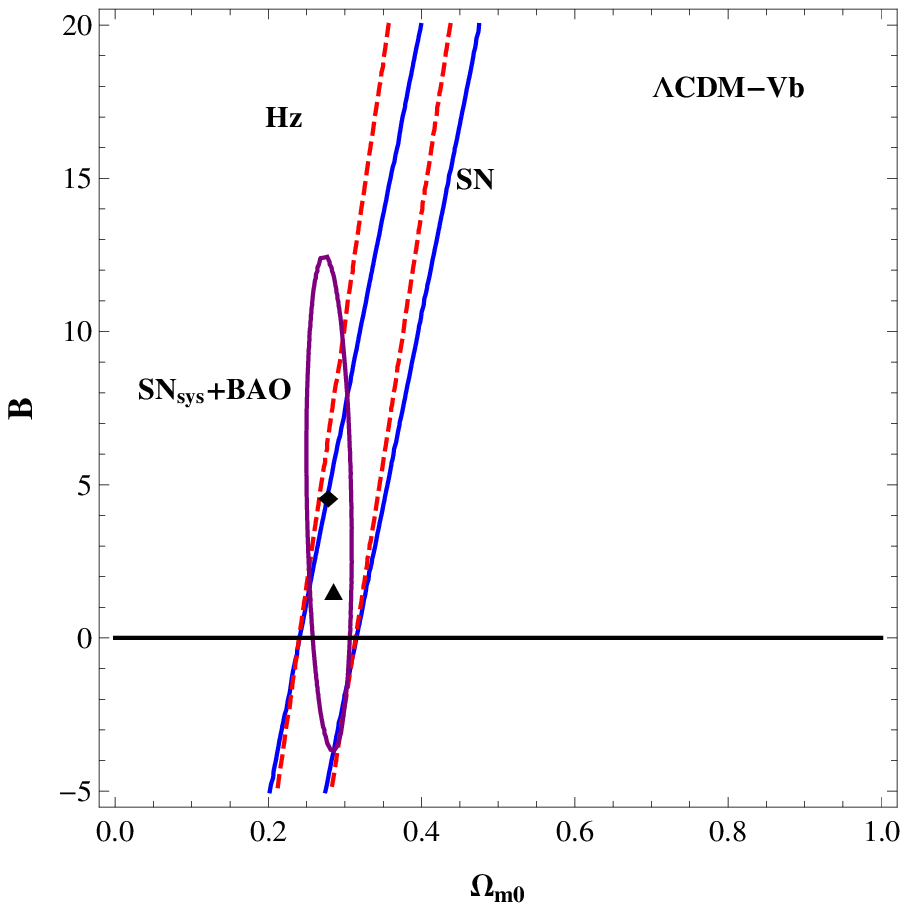}
\end{center}
\caption{$2 \sigma$ confidence ranges for parameter pair $(\Omega_{m0},B)$, constrained by SNe without systematic errors (blue), H(z) data-sets (purple) and joint analysis of BAO and SNe with systematic errors. The filled diamond and triangle represent best fit points given by SNe$+$BAO and H(z) respectively.}
\label{fig:vcr}
\end{figure}

The best-fit results for model Va and Vb are listed in Table~(\ref{tab1}), besides which the confidence ranges of parameter pair $(\Omega_{m0},B)$ of both models are shown in Fig.~(\ref{fig:vcr}).

\begin{table}[h!]
\caption{29 measurement points of the Observational Hubble Parameter Data-sets, we combined the data list given in~\cite{h0} with the latest data obtained from BOSS DR11~\cite{h8}.
}
\label{tab0}
\begin{tabular}{cccc}
\hline\noalign{\smallskip}
    redshift &  H $(km\cdot s^{-1}\cdot Mpc^{-1})$ & $\sigma_{\rm H}$ & Reference \\
\hline
0.100&	69&	12&	\cite{h1}\\
0.170&	83&	8&	\cite{h1}\\
0.270&	77&	14&	\cite{h1}\\
0.400&	95&	17&	\cite{h1}\\
0.900&	117&	23&	\cite{h1}\\
1.300&	168&	17&	\cite{h1}\\
1.430&	177&	18&	\cite{h1}\\
1.530&	140&	14&	\cite{h1}\\
1.750&	202&	40&	\cite{h1}\\
0.480&	97&	62&	\cite{h2}\\
0.880&	90&	40&	\cite{h2}\\
0.179&	75&	4&	\cite{h3}\\
0.199&	75&	5&	\cite{h3}\\
0.352&	83&	14&	\cite{h3}\\
0.593&	104&	13&	\cite{h3}\\
0.680&	92&	8&	\cite{h3}\\
0.781&	105&	12&	\cite{h3}\\
0.875&	125&	17&	\cite{h3}\\
1.037&	154&	20&	\cite{h3}\\
0.35&	76.3& 5.6&	\cite{h4}\\
0.07&	69.0&	19.6&		\cite{h5}\\
0.12&	68.6&	26.2&		\cite{h5}\\
0.20&	72.9&	29.6&		\cite{h5}\\
0.28&	88.8&	36.6&		\cite{h5}\\
0.44&	82.6&	7.8&	 	\cite{h6}\\
0.60&	87.9&	6.1&	 	\cite{h6}\\
0.73&	97.3&	7.0&	 	\cite{h6}\\
2.30&	224.0&	8.0&	\cite{h7}\\
2.36&	226&		8& \cite{h8}\\
\noalign{\smallskip}\hline
\end{tabular}
\end{table}

\begin{table}[h!]
\caption{6 measurement points of the Baryon Acoustic Oscillation Data-sets, we combined the data list given in~\cite{bao1} with new data from SDSS-III BOSS~\cite{bao2}.
}
\label{tab00}
\begin{tabular}{cccc}
\hline\noalign{\smallskip}
    redshift &  $\mathcal{A}$ & $\sigma_{\mathcal A}$ & Sample \\
\hline
0.106&	0.526&	0.028& 	6dFGS~\cite{bao1}\\
0.20&		0.488&	0.016&	SDSS~\cite{bao1}\\
0.35&		0.484&	0.016& 	SDSS~\cite{bao1}\\
0.44&		0.474&	0.034&	WiggleZ~\cite{bao1}\\
0.57&		0.436&	0.017&	BOSS~\cite{bao2}\\
0.6&		0.452&	0.018&	WiggleZ~\cite{bao1}\\
0.73&		0.424&	0.021& 	WiggleZ~\cite{bao1}\\
\noalign{\smallskip}\hline
\end{tabular}
\end{table}

\begin{table}[h!]
\caption{Best-fit parameters of $\Lambda$CDM-V models by various data-sets, where the reduced $\chi^2$ is chi-square divided by the degree of freedom ( which is $N-n-1$, where $N$ is the number of observations, and $n$ is the number of fitted parameters). The constraining results of joint analysis of H(z) and BAO shown here are not presented in Fig.~(\ref{fig:vcr}).}
\label{tab1}
\begin{tabular}{cccc}
\hline\noalign{\smallskip}
    data sets applied & H(z) & SNe(no sys-errs) & H(z)+BAO \\
\hline
model-Va\\
\hline
$reduced~\chi^2_{min}$ & 0.643 & 0.974 & 0.586 \\
$\Omega_{m0}$ & 0.277 & 0.283 & 0.278 \\
$B(=12\pi G\zeta_0T_0^\frac{1}{2})$ & -0.634 & 0.867 & 4.806 \\
$H_0$~~($km/s/Mpc$) & 68.293 & 69.884 & 70.656 \\
\hline
model-Vb\\
\hline
$reduced~\chi^2_{min}$ & 0.643 & 0.974 & 0.575 \\
$\Omega_{m0}$ & 0.286 & 0.290 & 0.280 \\
$B(=12\pi G\zeta_0T_0^\frac{3}{2})$ & 1.588 & 1.603 & 3.720 \\
$H_0$~~($km/s/Mpc$) & 68.585 & 69.885 & 70.955 \\
\noalign{\smallskip}\hline
\end{tabular}
\end{table}

The observational H(z) or SNe Ia data alone does not provide much tight constraints on parameter pair $(\Omega_{m0},B)$, until BAO data is taken into consideration. The overlapped $2\sigma$ confidence ranges in Fig.~(\ref{fig:vcr}) mildly favour positive value of $B$, and confine it under the magnitude of $20$.  The combined constraints for $\Omega_{m0}$ are also acceptable, which lie around $0.30$. The two panels in Fig.~(\ref{fig:vcr}) are similar to each other with tiny differences, indicating that the value of $\alpha$ is allowed between $1/2$ and $3/2$. The confidence ranges of $H_0$ are not shown here, since the constraints on the Hubble parameter by each individual data set are tight and give out similar values around $70~km/s/Mpc$. Besides, among the three parameters, the values of $B$ and $\Omega_{m0}$ matter most to our newly viscous model constraining.

\section{Discussions on the $\Lambda$CDM-V model}

The evolutional behaviour of viscosity is related to the expansion scalar $a=1/(1+z)$ which has been pointed out in the second section. The magnitude of viscosity will be gradually reduced by the on-going expansion. This provide us an physically acceptable transition mechanism of fluid from imperfectness to perfectness during the cosmic evolution.

According to the results given by model constraining, both model Va and Vb are acceptable. The results provide an allowed range for the value of $B$, and consequently set an upper limit for viscosity parameter. Since in this present study we are mainly interested in the evolutional effect of viscosity on late-stage cosmological evolution of the universe, in the following we will show how much deviation the viscosity model constrained by the background data-sets may provide against the standard scenario, the  $\Lambda$CDM model.

For convenience, we re-express the $\Lambda$CDM model in a simple form with the $\Omega_{m0}$ as the cosmic observational matter composite fraction with repeat to the critical density of the observable universe today
\begin{equation}
  {\rm H}_{\Lambda CDM}^2={\rm H}^2_0[\Omega_{m0}(1+z)^3+(1-\Omega_{m0})] .
\end{equation}
According to the latest WMAP-9yr observational results~\cite{wmap9y1}, we set the parameters $\Omega_{m0}=0.287$ and ${\rm H}_0=69.32$ by the global best-fit results for the standard cosmology model. In order to observe the effects of viscosity, we also set the parameter $A=0.427$ and ${\rm H}_0=70$ for model Va and Vb, while the value of parameter $B$ varies in $[0,20]$ for Va and $[0,10]$ for Vb  roughly according to the constraining results.

\begin{figure}[h!]
\begin{center}
\includegraphics[width=3in]{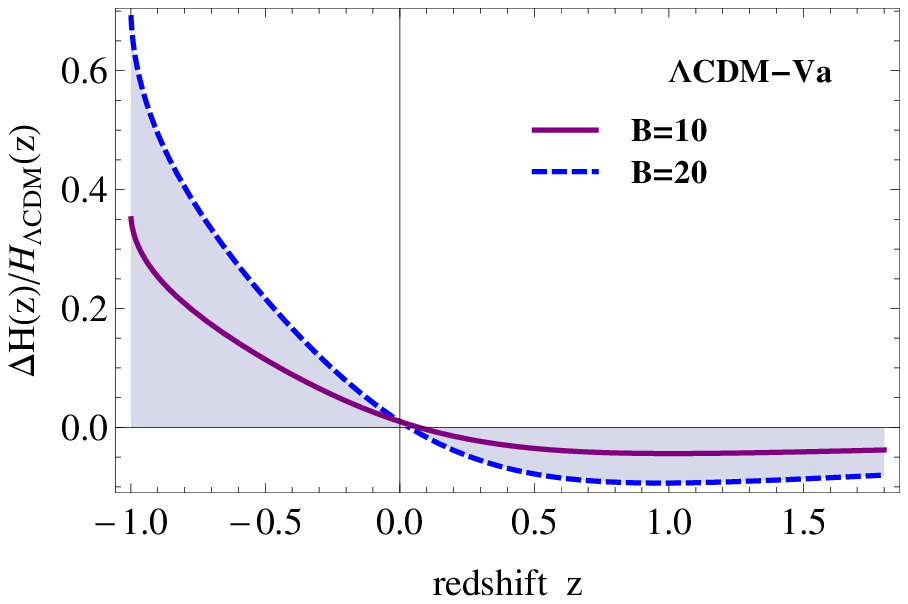}
\includegraphics[width=3in]{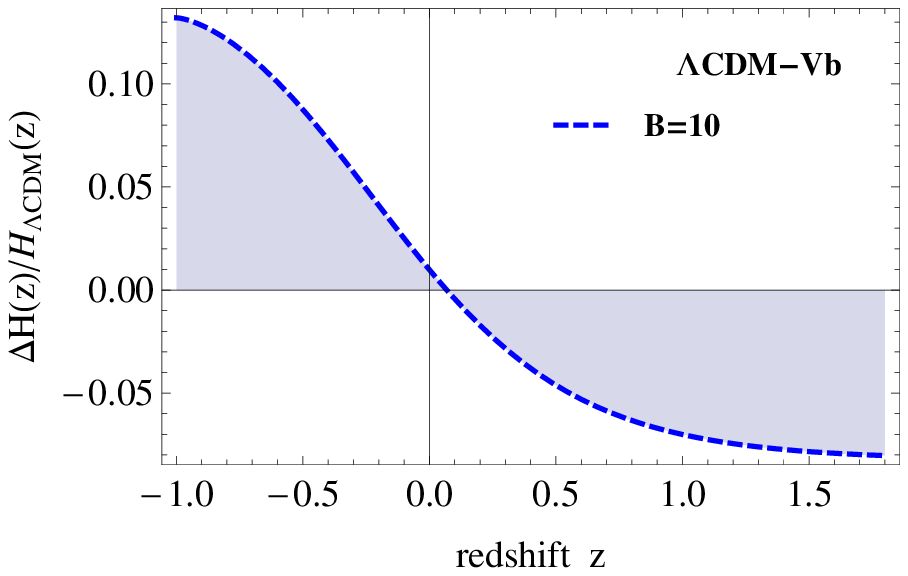}
\end{center}
\caption{ Ratio of deviation of H-z relation, where $\Delta {\rm H} = {\rm H}_{model}(z)-{\rm H}_{\Lambda CDM}(z)$, the shadowed area represents the possible H(z) values of the specific model with the value of B varying in the allowed range.}
\label{fig:vhz}
\end{figure}

The contrasts with H-z relation between the viscous models and standard cosmology models are illustrated in Fig.~(\ref{fig:vhz}). Viscosity contribution suppresses the value of Hubble parameter at high red-shift, which means that it increases the cosmic age. As we know today that the age of the universe is $13.772\pm0.059~Gyr$ as predicted by the standard cosmology model with observations~\cite{wmap9y1}. For the possible largest value of parameter $B$, the standard cosmic age could be extended by $1.012$ and $1.041$ times by Va and Vb model respectively. Generally speaking, the age of the universe predicted by the $\Lambda$CDM-V model will not be older than an upper limit as $15~Gyr$. This result is in consistent with the Planck data about cosmic age~\cite{planck}.

We are also interested in the fate of the universe predicted by the viscous model. To some extent, the $\Lambda$CDM-V model can be regard as a phantom-like dark energy model according to Eq.~(\ref{eq21}), (also see Fig.~(\ref{fig:wwp})).  As the expansion goes on the observational universe will tend to infinities, the cosmic media viscosity will fade out or be negligent totally and the imperfect fluid will gradually become perfect one to satisfying accuracy.

There are basically four types of future singularity~\cite{singularity}. Type I (Big Rip): the scale factor becomes infinite at a finite time in the evolution future. Type II (Sudden Rip): the scale factor and energy remain finite at the rip time when pressure becomes infinite. Type III: the scale factor remains finite when energy and pressure become infinite in the finite future. Type IV: the scale factor remains finite in the finite future when energy and pressure vanish, and the higher derivatives of H diverge. We may extend the classification to include the EoS singularity as well.

When the red-shift approaches $-1$, the Hubble parameter remains finite as shown in Fig.~(\ref{fig:vhz}), which means the Big Rip is avoided by model Va and Vb; according to Eqs.~(\ref{eq21}), effective pressure will always be finite, so Type II and III singularities are also absent. The last kind of singularity is apparently not predicted, so the $\Lambda$CDM-V model will not experience the above four types of singularity as well as the EoS singularity obviously.

Taking into account of the no-rip, little rip and pseudo-rip~\cite{rip} cases, there are seven types of fate for the universe evolution.  The $\Lambda$CDM model certainly predicts the no-rip future. We have already known that during the expansion of the universe, our viscous model will gradually evolve into the standard perfect one. With the Hubble parameter predicted to be constant (dominated by a cosmology constant like term) when time goes to infinity, the $\Lambda$CDM-V models are free of any kind of rip destiny.

\section{Conclusions}

In the above sections we have explored  the possible effects of the existence of temperature related viscosity on cosmological evolution by proper modelling which is supported by observed data-sets fittings. Similar research work in literature have proposed a variety of kinds of viscosity forms, from constant viscosity parameter~\cite{schwarz1,schwarz2} to complicated turbulence phenomena related forms as possibly  appearing in the complex cosmos media physics~\cite{turb}, and the temperature-dependent viscosity in this present work is proved to be one possibility with suitable models.

In this article, at first we have attempted to conceive the  unified dark energy (matter, dark matter and dark energy unified) models, which is rejected by the fact that it may cause singularity during cosmic evolution. Similar researches have pointed to basically the same result, which indicates that unified fluid models with viscosity can mimic the evolutional behaviour, but usually cause damping of density perturbations~\cite{schwarz1,JDB1,JDB2,JDB3,JDB4}.

Following the extension of our first trial we have then proposed the $\Lambda$CDM-V model, which is a modification of the standard $\Lambda$CDM model accordingly, turning the idealized perfect fluid cosmology model into the practically imperfect one with proper viscous energy-momentum tensor functioning and is fitted satisfactorily with the currently main observational data-sets.

\begin{figure}[h!]
\begin{center}
\includegraphics[width=3in]{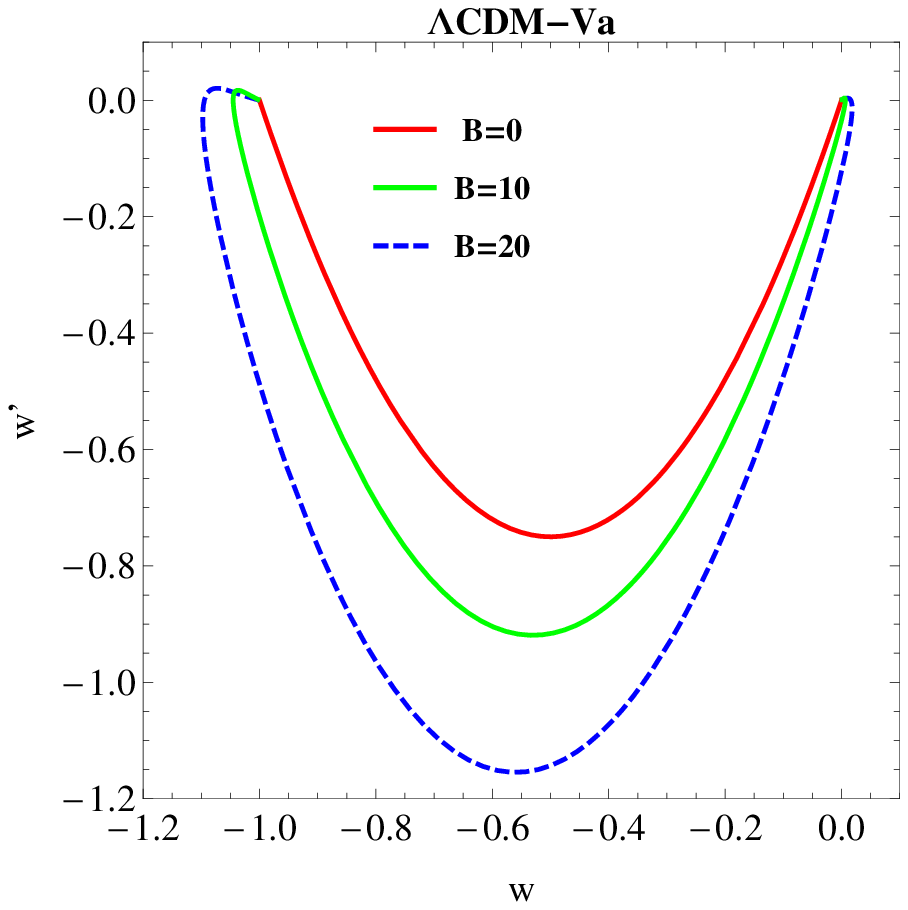}
\includegraphics[width=3in]{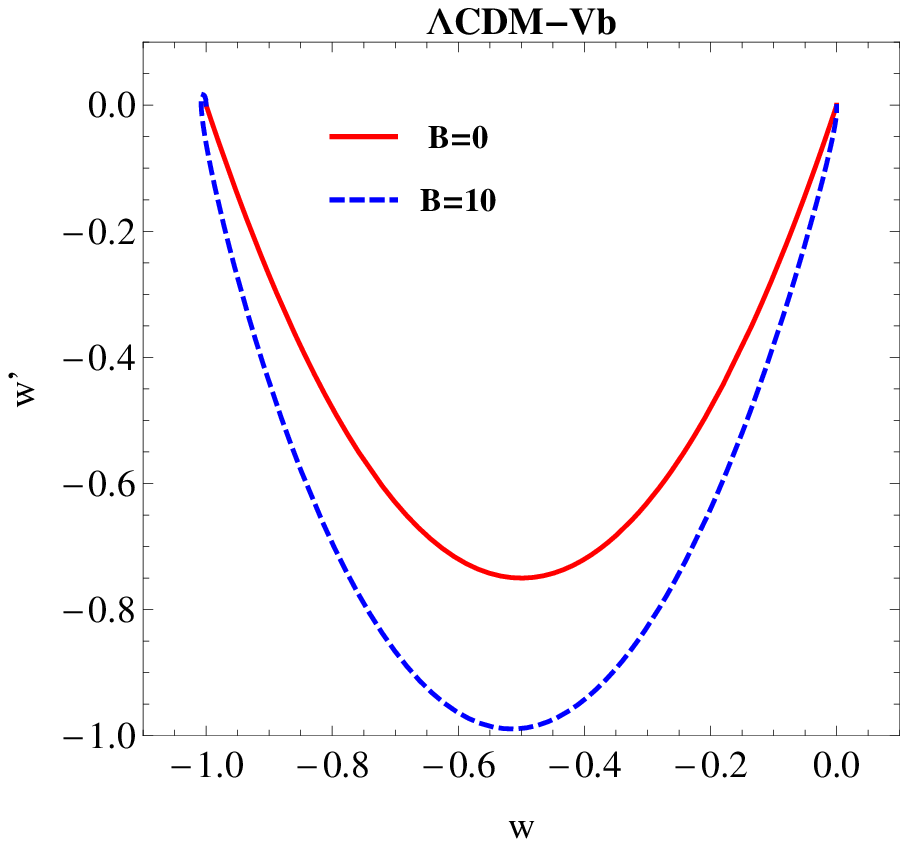}
\end{center}
\caption{The w-w' phase space of the $\Lambda$CDM-V models, where the prime represents $\frac{\partial}{\partial z}$ instead of the $\frac{\partial}{\partial N}$ with $N=\ln a(t)$ and $w$ here is the effective EOS parameter since we hope to give a clear behavior of the effective equation of state especially when $w=-1$ rather than  to illustrate the dark energy model classification by Caldwell and Linder parametrization (Ref.~\cite{wwp1,wwp2,wwp3}). The red line represents the w-w' curve of the $\Lambda$CDM model, while the dashed line shows the limit of viscous models in phantom region.}
\label{fig:wwp}
\end{figure}

In the Fig.~(\ref{fig:wwp}) for numerical comparisons, we can see the parametrized equation of state (EoS) parameter evolution lines of the $\Lambda$CDM-V and the standard $\Lambda$CDM model can match at present stage and in the future evolution, to form a consistent picture which is in accordance with the no-rip future predicted by the new model. In addition to increasing the cosmic age and being free of future singularity, we would like to say the smooth transition mechanism is another pleasing feature of the $\Lambda$CDM-V model. In this present work, the sub-model Va and Vb could converge to the cosmic perfect fluid model, the globally well fitted $\Lambda$CDM model, as the cosmic expansion goes on to infinity to dilute the effects of viscosity, thus provides a smooth transition mechanism from practically imperfect fluid models to perfect fluid ones. Fig.~(\ref{fig:wwp}) also indicate that our imperfect fluid model is phenomenologically related to phantom cosmology, since the EoS parameter of $\Lambda$CDM-V model is lower than $-1$. But in the far future, our model can predict a non-singular universe which can hardly be reached by phantom dark energy theories~\cite{phantom1,phantom2} or previous impefrect fluid models~\cite{example1,example2,example3,example4,example5,schwarz2,JDB1,JDB2,JDB3,JDB4}.

Considering about the complexities and difficulties for the physical viscous media descriptions in observational universe,  in the present work we have not included the density perturbation discussions which may give much tighter constraints on model buildings, like via the integrated Sachs-Wolfe effect or matter power spectrum simulations (see Refs.~\cite{schwarz1,schwarz2,velten1,velten2}). We strongly believe that more precise observational data-sets given by future astrophysics observation instrumental achievements can certain provide higher capacity performance in model constraining, constructing and will finally pin down the physics for long puzzling cosmic dark matter problems and mysterious dark energy phenomena.

The currently possible origins and explanations of viscosity physics may remind us of that trying to express viscosity effect in only one form could be too simple. The rich viscosity effects in cosmos may be an assemblage of a variety of dynamical and kinematics effects of locally cosmic motion and globally cosmos large-scale evolution, which may show different features when observed on different scales and/or in different evolutional stages. We may say present research reveals only a tip of the whole iceberg. Nevertheless, continuous study endeavours to the mysterious dark sector
physics will surely shed light towards the fundamental understandings of our universe.

\section*{Acknowledgement}
Interesting communications with Profs. S.~Odintsov and Lewis~H.~Ryder are very enjoyable. We thank Prof. J.~D.~Barrow for kindly pointing out Refs.~\cite{JDB1,JDB2,JDB3,JDB4} to us and Dr. Hermano~Velten for offering precious suggestions and help.

This work is partly supported by Natural Science Foundation of China under Grant Nos.11075078 and 10675062, and by the project of knowledge Innovation Program (PKIP) of Chinese Academy of Sciences (CAS) under the grant No.~KJCX2.YW.W10 through the KITPC astrophysics and cosmology program where we have initiated this present work.


\newpage

\begin{thebibliography}{99}
\bibitem{winberg}
D.~H.~Winberg {\it et al.}, arXiv:1201.2434v1 (2012).

\bibitem{wmap9y1}
G.~Hinshaw {\it et al.}, arXiv:1212.5226 (2012).

\bibitem{gbz}
Gong-Bo~Zhao, {\it et al.}, Phys.\ Rev.\ Lett. {\bf 109}, 171301 (2012).

\bibitem{xh1}
Xin-He~Meng, Zhi-Yuan~Ma, Eur.\ Phys.\ J.\ C {\bf 72}, 2053 (2012).
\bibitem{xh2}
Xin-He~Meng, Jie~Ren, Ming-Guang~Hu, Commun.\ Theor.\ Phys. {\bf 47}, 379-384 (2007).
\bibitem{xh3}
Xu~Dou, Xin-He~Meng, Adv.\ Astron. {\bf 2011}, 829340 (2011).

\bibitem{example1}
I.~Brevik {\it et al.}, Phys.\ Rev.\ D {\bf 86}, N6, 063007 (2012).
\bibitem{example2}
I.~Brevik {\it et al.}, Phys.\ Rev.\ D {\bf 84}, 103508 (2011).
\bibitem{example3}
T.~Padmanabhan, S.~Chitre, Phys.\ Lett.\ A {\bf 120} (1987) 433.
\bibitem{example4}
M.~Cataldo, {\it et al.}, Phys.\ Lett.\ B {\bf 619} (2005) 5.
\bibitem{example5}
Oliver~F.~Piattella {\it et al.}, JCAP, 1105 (2011), 029.

\bibitem{turb}
I.~Brevik, {\it et al.}, Phys.\ Rev.\ D {\bf 86}, 063007 (2012).

\bibitem{schwarz2}
H.~Velten, D.~J.~Schwarz, Phys.\ Rev.\ D {\bf 86}, 083501 (2012), arXiv:1206.0986v3.

\bibitem{JDB1}
B.~Li, J.~D.~Barrow, Phys.\ Rev.\ D {\bf 79} 103521 (2009).
\bibitem{JDB2}
J.~D.~Barrow, Phys.\ Lett.\ B {\bf 180}, 335 (1986).
\bibitem{JDB3}
J.~D.~Barrow, Nucl.\ Phys.\ B {\bf 310}, 743 (1988).
\bibitem{JDB4}
J.~D.~Barrow, String-driven Inflation, in \textit{The Formation and Evolution of Cosmic Strings}, eds.
G.~Gibbons, S.~W.~Hawking, T.~Vachaspati, pp 449-464, CUP, Cambridge (1990).

\bibitem{schwarz1}
H.~Velten, D.~J.~Schwarz, JCAP.\ {\bf 1109}, 016 (2011), arXiv:1107.1143v2.

\bibitem{cmbt1}
P.~Noterdaeme, {\it et al.}, arXiv:1012.3164.

\bibitem{cmbt2}
J.~M.~LoSecco, G.~J.~Mathews, Yun~Wang, Phys.\ Rev.\ D {\bf 64}, 123002 (2001).

\bibitem{singularity}
Shin'ichi~Nojiri, {\it et al.}, Phys.\ Rev.\ D {\bf 71}, 063004 (2005).

\bibitem{rip}
P.~H.~Frampton, {\it et al.}, Phys.\ Rev.\ D {\bf 85}, 083001 (2012).

\bibitem{phantom1}
Bohdan Novosyadlyj, {\it et al.}, Phys.\ Rev.\ D {\bf 86}, 083008 (2012).
\bibitem{phantom2}
Mauricio Cataldo, {\it et al.}, arXiv: 1302.3748.
\bibitem{wwp1}
R.~R.~Caldwell, E.~V.~Linder, Phys.\ Rev.\ Lett. {\bf 95}, 141301 (2005).
\bibitem{wwp2}
R.~R.~Caldwell, Phys.\ Lett.\ B {\bf 545} 23-29 (2002).
\bibitem{wwp3}
Y.~D.~Xu {\it et al.}, Astrophys.\ Space.\ Sci. {\bf 337}, 493-498 (2012).
\bibitem{planck}
Planck Collaboration, arXiv:1303.5076v1, Submitted to Astronomy \& Astrophysics.
\bibitem{velten1}
J.~C.~Fabris, H.~E.~S.~Velten, W.~Zimdahl, Phys.\ Rev.\ D {\bf 81}, 087303 (2010), arXiv:1001.4101v1.
\bibitem{velten2}
J.~C.~Fabris, H.~E.~S.~Velten, Phys.\ Lett.\ B {\bf 694}, 289-293 (2011), arXiv:1007.1011v1.
\bibitem{weihao}
Hao~Wei, JCAP. \ {\bf 1104}, 022 (2011), arXiv:1012.0883v3.
\bibitem{constraint}
S.~Capozziello,S.~Nojiri, S.~D.~Odintsov, {\it et al.}, Phys.\ Rev.\ D {\bf 73}, 043612 (2006) arXiv:astro-ph/0508350v3 (2006).
\bibitem{Du1}
Meng~Xin-He, Du~Xiao-Long, Commun.\ Theor.\ Phys.\ {\bf 57}, 227-233 (2012)
\bibitem{Du2}
Xinhe~Meng, Xiao-Long~Du, Phys.\ Lett.\ B {\bf 710}, 493-499 (2012)

\bibitem{sys-err}
O.~Farooq {\it et al.} Astrophys. J. {\bf 764}, 138 (2013)

\bibitem{h0}
O~.Farooq, Bharat Ratra, Astrophys. J. {\bf 766}, L7 (2013).
\bibitem{h1}
J.~Simon, L.~Verde, R.~Jimenez, Phys.\ Rev.\ D {\bf 71}, 123001 (2005).
\bibitem{h2}
D.~Stern, R.~Jimenez, L.~Verde, M.~Kamionkowski, S.~A.~ Standford, J.\ Cosmology Astropart.\ Phys, {\bf 02}, 008 (2010).
\bibitem{h3}
M.~Moresco, A.~Cimatti, R.~Jimenez, {\it et al.} J.\ Cosmology Astropart.\ Phys., arXiv:1201.3609v1. (2012).
\bibitem{h4}
C.~H.~Chuang, Y.~Wang, [accepted by Mon. Not. Roy. Astron. Soc.], arXiv:1209.0210.
\bibitem{h5}
Cong~Zhang, Han~Zhang, Shuo~Yuan, Tong-Jie~Zhang, Yan-Chun~Sun, arXiv:1207.4541 (2012).
\bibitem{h6}
C.~Blake,S.~Brough,M.~Colless {\it et al.},  Mon.\ Not.\ Roy.\ Astron.\ Soc., {\bf425}, 405, arXiv:1204.3674 (2012).
\bibitem{h7}
N.~G.~Busca {\it et al.}, Astron.\ Astro. {\bf 552} (2013) A96.
\bibitem{h8}
A.~Font-Ribera {\it et al.}, arXiv:1311.1767.

\bibitem{bao1}
C.~Blake {\it et al.}, Mon.\ Not.\ R.\ Astron.\ Soc. {\bf 418}, 1707 (2011).
\bibitem{bao2}
C.~H.~Chuang {\it et al.}, arXiv:1303.4486.

\bibitem{sdss}
Percival, W. J., Reid, B. A., Eisenstein, D. J., \textit{et al}., MNRAS \textbf{401}, 2148 (2010).
\bibitem{age}
Mi-Xiang~Lan, Miao~Li, {\it et al.}, Phys.\ Rev.\ D {\bf 82}, 023516 (2010) arXiv:1002.0978v3 (2010).
\bibitem{HnS}
W.~Hu and N.~Sugiyama, Astrophys.\ J. {\bf 471}, 542 (1996).


\bibitem{liddle}
Andrew~Liddle, \textit{An Introduction to Modern Cosmology 2ed} (Weiley, 2003).
\bibitem{Champan}
S.~Chapman, T.~G.~Cowling, \textit{The Mathematical Theory of Non-uniform Gases 3rd} (Cambridge, 1995).
\bibitem{Joseph}
J.~O.~Hirschfelder, C.~F.~Curtiss, R.~B.~Bird, \textit{Molecular Theory of Gases and Liquids} {Wiley, 1964}.
\bibitem{landau}
L.~D.~Landau, E.~M.~Lifshitz, \textit{Fluid Mechanics 2ed} (Elsevier, 2004).
\bibitem{Luca}
Luca~Amendola, Shinji~Tsujikawa, \textit{Dark Energy: Theory and Observations} (Cambridge, 2010)

\end{thebibliography}
\end{document}